\documentstyle[twocolumn,aps,prb,epsf]{revtex}

\begin{document}
 
\title{Prediction of Protein Secondary Structures From 
Conformational Biases}
 
\author{\bf Trinh Xuan Hoang$^{1,2}$, Marek Cieplak$^{3,4}$, 
Jayanth R. Banavar$^3$, and Amos Maritan$^{1,2}$}
 
\address{(1) International School for Advanced Studies (SISSA/ISAS)
and INFM, via Beirut 2-4, 34014 Trieste, Italy}

\address{(2) The Abdus Salam International Center for Theoretical
Physics (ICTP), Strada Costiera 11, 34100 Trieste, Italy}

\address{(3) 104 Davey Laboratory, The Pennsylvania State University,
University Park, Pennsylvania 16802}
 
\address{(4) Institute of Physics, Polish Academy of Sciences, 
02-668 Warsaw, Poland}

\maketitle

%\vskip 20pt
%\noindent
%$^*$Correspondence to: \\
%Trinh Xuan Hoang,\\
%The Abdus Salam International Center for Theoretical Physics (ICTP), \\
%Condensed Matter Group, \\
%Strada Costiera 11, 34100 Trieste, Italy \\
%Tel: +39-040-2240-460 \\
%Fax: +39-040-224163 \\
%E-mail: hoang@sissa.it

%\vskip 20pt
%\noindent
%Grant sponsors: INFM, NASA.
%%, and KBN (Poland) -  2P03B-146-18.

%\vskip 40pt

\newpage
\begin{abstract}
{
We use LINUS, a procedure developed by Srinivasan and Rose, to provide
a physical interpretation of and to predict
the secondary structures of proteins. The 
secondary structure type at a given site is identified by
the largest conformational bias during short time simulations.
We examine the rate of successful prediction as a function
of temperature and the interaction window. At
high temperatures,  there is a large propensity for the
establishment of $\beta$-strands whereas 
$\alpha$-helices appear only when the temperature
is lower than a certain threshold value. 
It is found that there exists 
an optimal temperature at which the correct
secondary structures are predicted most accurately.
We find that this temperature is
close to the peak temperature of the specific heat. 
Changing the interaction window or carrying out longer
simulations approaching equilibrium
lead to little change in the optimal success rate.
Our findings are in accord with the  observation by Srinivasan and Rose 
that the secondary structures are mainly determined by local
interactions and they appear in the early stage of folding.
}
\end{abstract}


\vskip 20pt

%\newpage
\section*{Introduction}
A knowledge of the three dimensional structure of a protein is crucial
for understanding its biological functionality. 
Unfortunately, the rate at which protein structures can be experimentally 
solved is far behind the speed at which the sequences are determined.
With progress in the Human Genome Project, 
a good computer-based method for the prediction of protein structures
from their sequences would be an invaluable tool for modern
microbiology as well as for drug design. The existing methods
for structure prediction can be divided into two classes:
1) template-based methods which compare a sequence with
unknown structure against the library of solved structures
and 2) ab initio methods which seek to identify the native fold usually
defined as the lowest energy point in conformational space. 
The latter are specially useful when a target sequence has a
low similarity with the existing protein sequences of known structures.
It should be noted though that many so called 
ab initio methods do use information derived
from the protein database as input.

Significant progress has been achieved in the ab initio approach
to protein structure prediction as witnessed in the CASP competitions,
\cite{Moult,Orengo,Baker_review} 
wherein the structures of large protein fragments, comprising as many as
100 residues, were predicted with an accuracy of 4-7 $\AA$ in rmsd.
A notable success reported was that of the Baker group and entailed the
assembly of protein conformations from fragments of known 
structures in the protein database, which have local sequences similar to 
that of the target sequence, using statistically derived scoring 
functions.\cite{Baker_casp3,Baker_jmb1997,Baker_jmb2001} 
In Levitt's approach,\cite{Levitt_jmb,Levitt_casp3} 
secondary structures, which were predicted
by using several existing secondary structure prediction methods,
\cite{Rost,Ross,Frishman}
are fitted to best scoring compact conformations obtained on a 
simplified tetrahedral lattice.
Scheraga and coworkers \cite{Scheraga_pnas,Scheraga_casp3}
use an off-lattice C$_\alpha$-based model
with interactions imposed on virtual side-chains and virtual
peptide groups. The lowest-energy C$_\alpha$ trace obtained by
extensive conformational space annealing is then converted to an all-atom
backbone for further refinements. 
Skolnick et al. \cite{Skolnick} 
built discretized protein conformations 
using predicted secondary structures and a number of tertiary restraints 
derived from multiple sequence alignments. 
The success of these ab initio methods relies to a large extent
on knowledge-based information, i.e. data  derived
from  known protein structures, such as that used in
the scoring functions, secondary structure prediction or
in the choice of fragments to incorporate in the model. 

Our work deals with secondary structure prediction and
builds on a truly ab initio protein structure prediction procedure 
called LINUS developed by Srinivasan and Rose.\cite{Rose1} 
LINUS does not use any knowledge-based information and thus provides
a clear picture of the role played by the different factors in folding.
Furthermore, the algorithm for determining the structure is not
based on energy minimization -- LINUS captures the interplay between 
energy and entropy in  determining the local secondary structure.

The most powerful aspect of LINUS is its simplicity -- it is based
on just 4 essential aspects of protein behavior:
(1) excluded volume, (2) preferred occupancies of the dihedral
angles in certain regions in the Ramachandran plot,\cite{Ramachandran} 
(3) hydrophobic interactions and hydrogen bonding and 
(4) the hierarchical organization of protein structures. 
\cite{Crippen,Rose_1979}
In spite of this simplicity, LINUS has already proved to be effective
in predicting the secondary and super-secondary structures of 
protein fragments. \cite{Rose1}
Note that the hierarchical algorithm steers folding along some
specific pathways and the resulting structure does not necessarily 
correspond to the global energy minimum. 

In a subsequent study,\cite{Rose2} 
Srinivasan and Rose used LINUS to propose a physical
basis for secondary structures,  
which showed that protein secondary structures are mainly 
determined by steric effects and local interactions.
This conclusion recently obtained strong support from experimental 
evidence that unfolded protein conformations, 
under highly denaturing conditions and thus in the absence of
long-range contacts, are still characterized by
local native-like topology. \cite{Shortle,Plaxco}
In LINUS, the conformational bias towards a type of secondary structure
is determined through the probability of being in this conformation 
during simulation.

We have found this idea to be intriguing and worthy of a careful reexamination.
Here, we make an assessment of how well the secondary structures can
be predicted based on the analysis of conformational biases.
In particular we concentrate on the role played by the temperature, $T$,  
in determining the success rate and find that there is an optimal $T$
at which the secondary structure prediction is the best. 
For most of the proteins studied, this temperature coincides
with the one that Rose and Srinivasan used
in their studies and is found to be  near the peak in the specific heat
where the conformational conversion in the system is the largest.
The optimal conditions for the structure prediction do not 
depend much on whether the window in the interactions
allowed for purely local or also for non-local interactions.
They are also insensitive
to the duration of the simulations. We obtained very similar
results when long, nearly equilibrium, simulations were considered.

The aim of our study is to elucidate how LINUS works and what its strengths
and weaknesses are. The ultimate goal would be to determine what kinds of
improvements could be made in this physically appealing framework to move
towards first principles tertiary structure prediction. 

\section*{Methods}

A detailed description of LINUS can be found in the original papers
of Srinivasan and Rose.\cite{Rose1,Rose2} 
We have developed our own version of LINUS that
strictly follows the improved development as described in the 
PNAS paper\cite{Rose2}.
Briefly, in LINUS, the
coordinates of all backbone atoms are considered whereas 
a sidechain is represented in a simplified manner.
Specifically, glycine has no sidechain, alanine's sidechain
is made of a C$_\beta$ and the remaining amino acids are
represented by C$_\beta$ and one or two pseudo C$_\gamma$
atoms, depending on whether the sidechain is branched out or not.
The atoms are modeled as hard spheres that are not allowed to overlap.
The sizes of the spheres depend on the type of the atom and
the sizes of the pseudo-atoms depend on the size of the
sidechains that they represent. 

Apart from steric interactions, the Hamiltonian consists of just a few terms
that provide attraction between atoms: hydrogen bonding (H-bond),
hydrophobic interaction, and salt bridges. 
All backbone nitrogens, except for those that belong to a proline,
are considered to be H-bond donors and participate in
no more than one H-bond but the
nitrogen at the N-terminus may participate in up to three H-bonds. 
The backbone oxygens and the sidechains of some
amino acids (Ser, Thr, Asn, Asp, Gln, Glu) are acceptors.
A backbone-to-backbone hydrogen bond is assumed to be formed 
between residues $i$ and $j$ when they are at least three residue
apart in the sequence and when the distance between a donor and 
an acceptor is smaller than 5$\AA$.
An energy of $-0.5\epsilon$ is assigned,
where $\epsilon$ is an energy unit,
and the energy is scaled quasi-linearly from 0 to its minimal 
value as the distance decreases to 3.5$\AA$. 
It is also required that the out-of-plane dihedral angle
O$(j)$--N$(i)$--C$_\alpha(i)$--C$(i-1)$ should be larger than $140^o$.
A sidechain-to-backbone hydrogen bond is formed when the donor-to-acceptor
distance is smaller than 4$\AA$ and the acceptor must be not further
than four residues away from the donor in the sequence.
In this case an energy of $-1.0\epsilon$ is
assigned and no scaling of the energy is involved.

Hydrophobic attraction is postulated to occur for contacts
between the sidechain atoms of hydrophobic (Cys, Ile, Leu, Met, Phe,
Trp, Val) and amphipathic (Ala, His, Thr, Tyr) residues. 
The minimal value of the contact energy is $-0.5\epsilon$ when both residues
are hydrophobic and $-0.25\epsilon$ when one of them is hydrophobic
and the other is amphipathic.
A contact  between two atoms $i$ and $j$ is said to form when the distance
between them is smaller than $R(i) + R(j) + 1.4\AA$, where
$R(i)$ and $R(j)$ are the contact radii of the two atoms.
The contact radii of the atoms \cite{Rose2} depend on the kind of atoms
and are larger than their hard sphere radii. 
The energy of a contact scales from 0 to its minimal value
as the distance between two atoms decreases from its cut-off value
to $R(i) + R(j)$. 
A salt bridge is assigned to contacts between oppositely charged
groups (namely the sidechains of Arg or Lys with Glu or Asp).  
The minimal energy of a salt bridge is $-0.5\epsilon$.
In LINUS there is also an energy function to chase residues away from the
right hand side of the Ramachandran plot. When a residue
has a positive torsional angle $\phi$ it is punished with an energy
of $-1.0\epsilon$ if the residue is not a glycine, otherwise it
is rewarded with an energy of $-1.0\epsilon$.

The main degrees of freedom used in LINUS are the Ramachandran
torsional angles $\phi$ and $\psi$ and the torsional $\chi$
which corresponds to rotation of the sidechains. 
Additionally the torsional angle $\omega$ about the peptide bond and
the N-C$_\alpha$-C bond angle are allowed to be perturbed slightly
during the simulation. All the other bond angles and bond
lengths are kept fixed.
Three consecutive residues $(i,i+1,i+2)$ are perturbed at a time 
and the movements advance from the N-terminus to the C-terminus.
The moves at an $i$th residue are repeatedly chosen until a move
is obtained in which 
there are no steric clashes within the three residue fragment
considered. At the next stage the whole protein chain
is checked for the presence of steric clashes.
Up to 50 such attempts are performed in order to find a conformation
without any steric clashes. If the new conformation found 
still has steric clashes it is rejected, otherwise it is accepted 
with a probability 
${\cal P} = min\left\{1,e^{-\Delta E/k_B T}\right\}$, where 
$k_B$ is the Boltzmann constant, $T$ is
the temperature measured in the units of $\epsilon/k_B$,
and $\Delta E$ is the energy difference. 
A complete progression from N to C is called a cycle.

LINUS uses a smart move set that consists
of the following, equally probable, move types
\begin{description}
\item[1. $\alpha$-helix]: three consecutive residues $(i-1,i,i+1)$
are set to having $\phi = -64 \pm 7^o$, $\psi = -43 \pm 7^o$.
\item[2. $\beta$-strand]: three residues $(i-1,i,i+1)$
are set to having $\phi = -130 \pm 15^o$, $\psi = 135 \pm 15^o$.
If a residue is a proline then $\phi$ is reset to $70\pm15^o$.
\item[3. turn]: there are 4 types of turns, namely I, I', II and II'.
For each turn type there are two possibilities:
a) setting residues $(i-1,i)$ to have turn $\phi$ and $\psi$
values while residue $i+1$ is set to random coil and b) setting
$i-1$ to random coil and  $(i,i+1)$ to have turn $\phi$ and $\psi$ values.
Overall there are 8 such possibilities. 
The turn $\phi$ and $\psi$ values for two consecutive residues
are given below for each type of a turn move (the notations used
for the residues are $i-1$ and $i$ but they can also be $i$ and $i+1$).
\begin{description}
\item [i.]Type I:\\
residue $(i-1)$:  $\phi=-60 \pm 15^o$, $\psi = -30 \pm 15^o$ \\
residue $(i)$: $\phi=-90 \pm 15^o$, $\psi = 0 \pm 15^o$ 
\item [ii.]Type I':\\
residue $(i-1)$:  $\phi=55 \pm 15^o$, $\psi = 40 \pm 15^o$ \\
residue $(i)$: $\phi=80 \pm 15^o$, $\psi = 5 \pm 15^o$ 
\item [iii.]Type II:\\
residue $(i-1)$: $\phi=-60 \pm 15^o$, $\psi = 110 \pm 15^o$ \\
residue $(i)$: $\phi=90 \pm 15^o$, $\psi = -5 \pm 15^o$ 
\item [iv.]Type II':\\
residue $(i-1)$: $\phi=60 \pm 15^o$, $\psi = -120 \pm 15^o$ \\
residue $(i)$: $\phi=-80 \pm 15^o$, $\psi = 0 \pm 15^o$ 
\end{description}
For all turn moves if a residue is a proline then its $\phi$ 
is reset to $70\pm15^o$.
\item[4. random coil]: $\phi$ and $\psi$ are chosen
randomly in one of the favorite regions of the Ramachandran plot. 
For non-glycine and non-proline residues
$(\phi,\psi) \in \{(-135 \pm 45^o,135\pm 45^o),
(-75 \pm 30^o,-30\pm 30^o), (75 \pm 15^o,30\pm 15^o)\}$. 
For glycine $\phi \in \{90\pm 30^o,180\pm 30^o\}$ and
$\psi \in \{0\pm30^o,180\pm30^o\}$. 
For proline $\phi=-70\pm15^o$ and $\psi \in \{135\pm30^o,-45\pm30^o\}$.

\end{description}
For the first three move types,
$\omega = 180 \pm 5^o$ whereas for the coil move $\omega = 180 \pm 10^o$. 
For all move types, the sidechain torsional angles ($\chi$s) are chosen at 
random in $10^o$ windows around $-60^o$, $60^o$ and $180^o$.

The conformational bias, $P$, of a given type of secondary structure
is defined as the probability of being in this structure
during the simulation. $P$ is usually computed as a function of residue
in the sequence.  The computation of $P$ requires a procedure
of secondary structure assignment, which allows one to determine 
which type of secondary motifs a residue belongs 
to at a given instant.
We use an assignment procedure in the most recent unpublished
development of LINUS,\cite{private}  
which proceeds through the following steps:
\begin{description}
\item[1.] Set all residues to the coil conformation ($c$).
\item[2.] For $i$ running from 1 through $N-3$, where $N$ is the number 
of residues, compute the torsion $\Theta$ between four consecutive 
C$\alpha$'s $(i,i+1,i+2,i+3)$.
\begin{description}
\item[a.] If $|\Theta| \geq 135^o$ then residues $(i+1)$ and $(i+2)$ are
set to the strand conformation ($s$).
\item[b.] If $45^o \leq \Theta \leq 65^o$ then residues $(i+1)$ and $(i+2)$ are
set to the helix conformation ($h$).
\item[c.] If $-50^o \leq \Theta < 45^o$ then residue $(i+1)$ 
and $(i+2)$ are set to the turn conformation ($t$).
\end{description}
\item[3.] Check again all residues from 1 through $N$:
\begin{description}
\item[a.] If a segment of less than 5 residues with a $h$ assignment
is found then all residues in this segment are set to $t$.
\item[b.] If a segment of less then 3 residues with a $s$ assignment
is found then all residues in this segment are set to $c$.
\end{description}
\end{description}

To compute $P$, one starts from an open conformation and
makes a simulation of 1000 cycles. After each cycle a conformation
assignment is determined to gather statistics on $P$. The average is taken
over 10 simulations for each $T$.

In order to make comparisons with the DSSP-based native assignments \cite{DSSP}
used in the PDB,\cite{PDB} we adopt a simplified correspondence in which
the $3_{10}$, $\pi$ and $\alpha$- helix correspond to $h$,
the isolated $\beta$-bridges and extended $\beta$-strands to $s$,
the hydrogen bonded turn to $t$, and bends and undefined segments
to $c$. It should be noted that the native state secondary
structure assignment used by Srinivasan and Rose \cite{Rose2}
for the proteins studied does not fully
agree with the one used in the PDB. In the following, our results are benchmarked
against the PDB-based assignment.

In order to explore the role of local and non local interactions 
we consider two choices for the interaction window, $\Delta$,
of 6 and $N$. 
The interaction window restricts interactions along
the sequence. $\Delta=6$ means that all interactions between
two residues $i$ and $j$ with $|i-j|>6$ are switched off, whereas
in the case of $\Delta=N$ all interactions are present.

\section*{Results and Discussion}

We begin our discussions with protein G (PDB code 1GB1), the protein
showing the best conformational biases towards native secondary
structures in the set of proteins studied by Srinivasan and Rose. 
Figure 1 shows $P$ as a function of residue at three different temperatures
$T=0.8\epsilon/k_B$, $0.5\epsilon/k_B$ and $0.2\epsilon/k_B$. 
$T=0.5\epsilon/k_B$ 
is the temperature which Srinivasan
and Rose used in their simulations. The interaction window is set to 6.
The conformational biases towards
$\alpha$-helices ($h$), $\beta$-strands ($s$), turns ($t$) and 
coils ($c$) are shown. 
Note that at $T=0.8\epsilon/k_B$ the strands dominate over all 
other structures. Thus the whole protein chain
prefers to be in the strand conformation at high temperatures. 
This follows from the simple observation that the entropy is largest
in the strand conformation and is the dominant factor in the free energy
at high temperatures.
At low temperatures, such as $T=0.2\epsilon/k_B$, 
the bias towards the strands 
vanishes while the highest biases belong to  
helices and turns.   This is because helices and turns involve favorable
interactions, which are predominantly local and are 
thus stabilized at low temperatures.
At the intermediate temperature, $T=0.5\epsilon/k_B$,
the dominating structure varies as
one proceeds along the sequence. Some parts of the protein
prefer to be in a strand conformation while others form 
helices and turns. 

Because the biases strongly depend on $T$, one may ask what the
temperature is at which the native secondary structure can be 
most reliably predicted.
In order to answer this question we have carried out an extensive analysis
of the biases over a wide range of temperatures.
Figure 2a shows two sequences of secondary structure assignments.
The first corresponds to the known native conformation of protein G,
and the second is obtained from the biases given in the middle
panel of Figure 1, i.e. at $T=0.5\epsilon/k_B$. In the latter case
an assignment at a given site is set to the type of secondary
structure showing the highest bias.
We introduce a parameter $\eta$ which estimates overlaps
between the two sets of assignments for each kind of secondary structure.
For a given type of conformation, $x$ ($x \in \{s,h,t,c\}$), 
$\eta$ is defined as the number of sites at which both assignments 
(from PDB and from the biases) are $x$ divided
by the number of sites at which at least one of the assignments is $x$.
Specifically if $A$ is a set of sites of type $x$ in the PDB assignment
and $B$ is a set of sites of the same type of conformation
predicted by the biases then 
\begin{equation}
\eta \equiv \frac{f(A \cap B)}{f(A \cup B)}\;\;\;,
\end{equation} 
where $f(X)$ is a function which returns the number of elements in $X$.
Thus, if $A \equiv B$ then $\eta=1$. We call $\eta$ the rate of successful
prediction.

Figure 2b shows $\eta$ as function of $T$ for the strands,
helices and turns for protein G. 
Note that the values of $\eta$ are the largest around $T=0.5\epsilon/k_B$.
At this temperature
the rate of prediction is the highest for helices and exceeds 90\%
(it is 100\% at $T=0.55\epsilon/k_B$), 
while strands and turns are predicted at 74\% and 23\% levels
respectively. The values of $\eta$ were obtained with the reference
to PDB assignment. If the Srinivasan and Rose assignment is
used instead, the corresponding success rates are 90\%, 63\%
and 35\% respectively.
As $T$ increases the rate for the strands
first decreases  and then remains roughly at a constant value while
the rate for helices drops rapidly and vanishes at $T=0.7\epsilon/k_B$. 
A nearly opposite scenario is observed as $T$ becomes smaller
than 0.5 -- the rate for strands drops rapidly while 
it remains high for the helices.

Figure 3 shows $\eta$ as a function of $T$ for protein G but
as calculated with $\Delta=N$. We still observe the same picture
as for $\Delta=6$, except that at low temperatures the
prediction rates for helices and strands become somewhat higher.
The optimal temperature, however, remains close to $0.5\epsilon/k_B$.

Figures 4 and 5 show $\eta$ as functions of $T$ for 6 other  proteins
in the set studied by Rose for $\Delta=6$ and $N$ respectively.
For $\Delta=6$, as in protein G,
the best prediction rates are obtained 
at about $T=0.5\epsilon/k_B$ 
for all of the proteins except for plastocyanin
(6PCY) and myo hemerythrin (2HMQ). The latter proteins are special
because they consist of only one kind of secondary structure in addition
to the turn.
The native conformation of plastocyanin is built only of
$\beta$-strands and that of hemerythrin is a four-helix bundle.
The best prediction rates are obtained for a range
of temperatures which corresponds to $T \ge 0.4\epsilon/k_B$ for 
plastocyanin and to $T \le 0.4\epsilon/k_B$ for hemerythrin. 
For $\Delta=N$, 
the optimal temperature varies a little but for most of the 
proteins it remains in the range from 0.4$\epsilon/k_B$ to 
0.6$\epsilon/k_B$.
It should also be noted that, when $\Delta = N$, the rates
for the strands at low temperatures become significantly larger
than in the $\Delta=6$ case for all proteins. 
The reason for this behavior is
that, at low $T$, the  strands can be stabilized only by non-local
interactions, which are absent when $\Delta=6$.
The results in Figures 2 through 5 are generally similar even when
the Srinivasan-Rose secondary structure assignment is used underscoring
the robustness of our results.  The only difference is that the predictions
pertaining to the turns are improved compared to the PDB-based secondary
structure assignment.

What is the principle that governs the choice of the optimal
temperature? Figure 6 shows how conformational changes occur
with respect to temperature for each residue of protein G 
%as the temperature decreases from 1.0$\epsilon/k_B$ to 0 
for the case of $\Delta=6$. 
It is seen clearly that most of the strands
are destabilized at low temperatures 
whereas helices are absent at high temperatures. 
Thus in order to have both kinds of structures predicted
the optimal temperature for the prediction should
be in a range of intermediate temperatures where helices have 
started to form but strands have not vanished. 
It can be seen in Figure 6 that as the temperature is lowered,
the strands undergo a transition to helices or other
kinds of structures such as turns or coils. A helix 
can also be formed from a coil as the temperature continues 
to decrease.
Because helices and turns are associated with the establishment of 
H-bonds while strands and coils usually have no contacts such transitions
entail a change in energy which is reflected in the specific heat.
This suggests that the optimal temperature for secondary structure prediction
ought to be in the vicinity of the peak of the specific heat, and likely
a bit higher than the temperature of its maximum 
in order not to discriminate against the strands. 

The connection between the thermodynamics of the system
and the optimal temperature for the prediction are shown in
Figures 7 and 8 for protein G and plastocyanin and 
hemerythrin respectively. 
The specific heat $C$ as a function of temperature is calculated
using the histogram technique.\cite{Swendsen}
For each protein we performed a long simulation of $200\,000$ cycles 
at $T=0.5\epsilon/k_B$ to deduce the thermodynamic behavior at that temperature
and other temperatures in its vicinity.
The results are shown for the two values of $\Delta$. For $\Delta=6$
the maximum in $C$ occurs roughly at $T=0.4\epsilon/k_B$ for the three proteins.
For $\Delta=N$ the magnitude of the 
peak in $C$ is higher and it also occurs at a slightly 
higher temperature due to the presence of the long range interactions.
(It is interesting to note that the experimentally determined specific
heat of plastocyanin \cite{heat} shows a sharp maximum around $68^o$C.)
Note that for protein G the optimal 
temperature for the secondary structure prediction is found 
in the vicinity of the peak in the specific heat -- just to the right
of the maximum.  
A similar behavior is observed in the case of plastocyanin 
except that the optimal temperature at $\Delta=6$ is 
farther from the maximum in $C$. This is due to the
fact that plastocyanin consists only of $\beta$-sheets and
the strands are more favored at high temperatures.
However, in the case of myo hemerythrin, whose native state contains
mainly $\alpha$-helices, the behavior is just the opposite.
The optimal temperature for the prediction is now
on the low temperature side of the peak in the specific heat.
A comparison of
Figures 4 and 5 shows that at the temperature corresponding to the
maximum in $C$, the rates of successful prediction are
already close to their best values. 

The results described so far are based on short
time simulations which last for 1000 cycles at each $T$.
Figure 9 shows $\eta$ as function of $T$ for protein G 
when the conformational biases are determined 
from nearly equilibrium simulations. 
These simulations are performed similar to the calculation of
the specific heat:
we make a long simulation of $200\;000$ cycles at $T=0.5\epsilon/k_B$
and then use the histogram method to obtain the biases
at other temperatures. The profiles of $\eta$ over $T$ are 
surprisingly similar to those shown in Figure 3 and the peaks
seem to be even more pronounced. The predictions given by
the conformational biases are found to be  insensitive to the length
of simulations, at least for a range of temperatures
which are close to the optimal value.

\section*{Conclusions}

Our results confirm that the 
analysis of  conformational biases is
a fast and useful tool to get information about native
protein secondary structures. 
We find that the most common secondary motifs,
$\alpha$-helices and $\beta$-strands,
can be predicted with an accuracy ranging from roughly 40\% up to 100\%.
Our analysis shows that while the rate of successful prediction
is insensitive to the interaction window as well as to 
the length of the simulations,
the choice of temperature appears to be critical.
The optimal run temperature is found to be related to the peak temperature 
in the specific heat.
Unlike commonly used algorithms in which one attempts to minimize
an energy function to determine the native state structure, LINUS
is an algorithm that relies on a delicate interplay between the entropy
favoring the strands and energetic considerations favoring turns and helices.
Because there is no procedure in LINUS that allows for an assembly of strands
through the appropriate non-local interactions into a sheet, secondary
structure prediction in essence depends on the persistence of a strand
conformation down to intermediate temperatures in regions corresponding to
strands in the native structure, while other regions adopt the helix and the
turn conformations due to the energy gain through the local contacts.  An
improvement in the prediction might be expected  on extending the "Local
Independently Nucleated Units of Structure" to some judiciously chosen
non-local interactions for the assembly of $\beta$-sheets.

\section*{Acknowledgments}
We are indebted to Raj Srinivasan and George Rose for inspiring
this work, giving us
the source code of LINUS and for providing us wonderful support
in understanding the implementation of LINUS.  This work was supported
by NASA and INFM.

\newpage
\centerline{FIGURE CAPTIONS}
\begin{description}
\item[Fig. 1.]
Conformational biases towards secondary structures, $P$, 
as functions
of residues determined for protein G at three different temperatures
$T=0.8\epsilon/k_B$, $0.5\epsilon/k_B$ and $0.2\epsilon/k_B$. 
The types of secondary structures are denoted as $h$
for $\alpha$-helices (continuous line), $s$ for $\beta$-strands (dotted line),
$t$ for turns (dashed line) and $c$ for coils (long-dashed line).
The biases are computed by averaging over 10 trajectories each of 
1000 cycles starting from an open conformation.
The interaction window is set to 6.
\item[Fig. 2.]
a)  The assignment of the secondary structures
for protein G extracted from the PDB structure using the
DSSP method of Kabsch and Sander\cite{DSSP}
(box) and predicted from the analysis of the conformational 
biases at $T=0.5\epsilon/k_B$ and for $\Delta=6$.
b) The rate of success in prediction, $\eta$, as a function of
temperature for three kinds of secondary structures: helix ($h$), strand ($s$)
and turn ($t$). The simulations are performed for protein G with 
$\Delta=6$. 
At each temperature studied, the conformational biases are computed by 
averaging over 10 trajectories each of 1000 cycles starting 
from an open conformation. 
The error bars are determined from three simulations at each 
temperature.
\item[Fig. 3.]
The rate of success in prediction, $\eta$, as a function of temperature for
three kinds of secondary structures: helix ($h$), strand ($s$) and turn ($t$)
for protein G with $\Delta = N$ or with no restriction on the range of
interactions.  The details are the same as in the lower part of Figure
\ref{sgb2}.
\item[Fig. 4.]
The rate of successful prediction of secondary structures
as a function of temperature for plastocyanin (6PCY), myo hemerythrin (2HMQ),
staphylococcal nuclease (1STG), ubiquitin (1UBQ), ribonuclease A (7RSA)
and ribonuclease H (2RN2).
The simulations are performed in the same way as for
protein G as described in the caption of Figure \ref{sgb2}. The interaction
window is set equal to 6.
\item[Fig. 5.]
Same as Figure \ref{eta} but with $\Delta=N$.
\item[Fig. 6.]
Top: Conformational diagram plotted as a function of temperature
and residue in protein G. 
The dark and light grey areas correspond to the helix ($h$)
and strand ($s$) conformations respectively.
In these calculations, $\Delta=6$.
Bottom: strand (thin box) and helix (thick box) fragments
found in the native conformation of protein G.
\item[Fig. 7.]
The specific heat as a function of temperature for protein G.
The thermodynamic averages were carried out by performing a long simulation
of $200\,000$ cycles at $T=0.5\epsilon/k_B$ and then using the histogram
method to extract quantities at other temperatures. The lower peak
(continuous line) and the higher one (dashed line) correspond to
the interaction window equal to 6 and $N$ respectively. The arrows
show the temperatures at which the secondary structures are best
predicted for the two values of $\Delta$.
\item[Fig. 8.]
Same as Figure \ref{cvgb1} but for plastocyanin (top) 
and myo hemerythrin (bottom).
\item[Fig. 9.]
Same as Figure \ref{agb2} but the rates of successful prediction are
computed from an equilibrium simulation. The secondary structure biases 
as a function of temperature are computed using the histogram method.
A long simulation of $200\,000$ cycles at $T=0.5\epsilon/k_B$ is performed to
extract quantities at other nearby temperatures.
\end{description}

%\newpage

% FIGURE 1
\begin{figure}
\epsfxsize=3.2in
\centerline{\epsffile{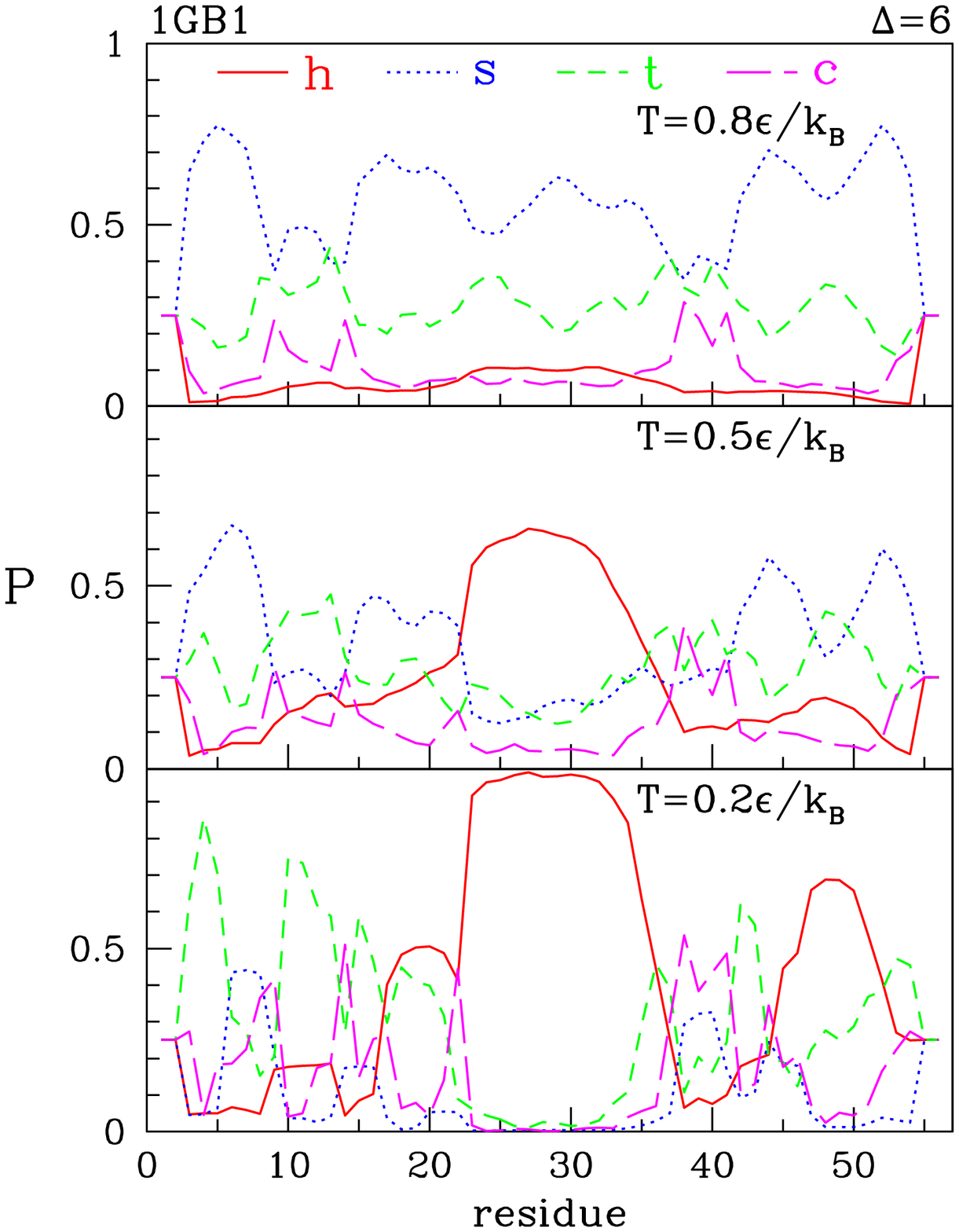}}
\caption{ }
\label{stgb2}
\end{figure}

% FIGURE 2
\begin{figure}
\epsfxsize=3.2in
\centerline{\epsffile{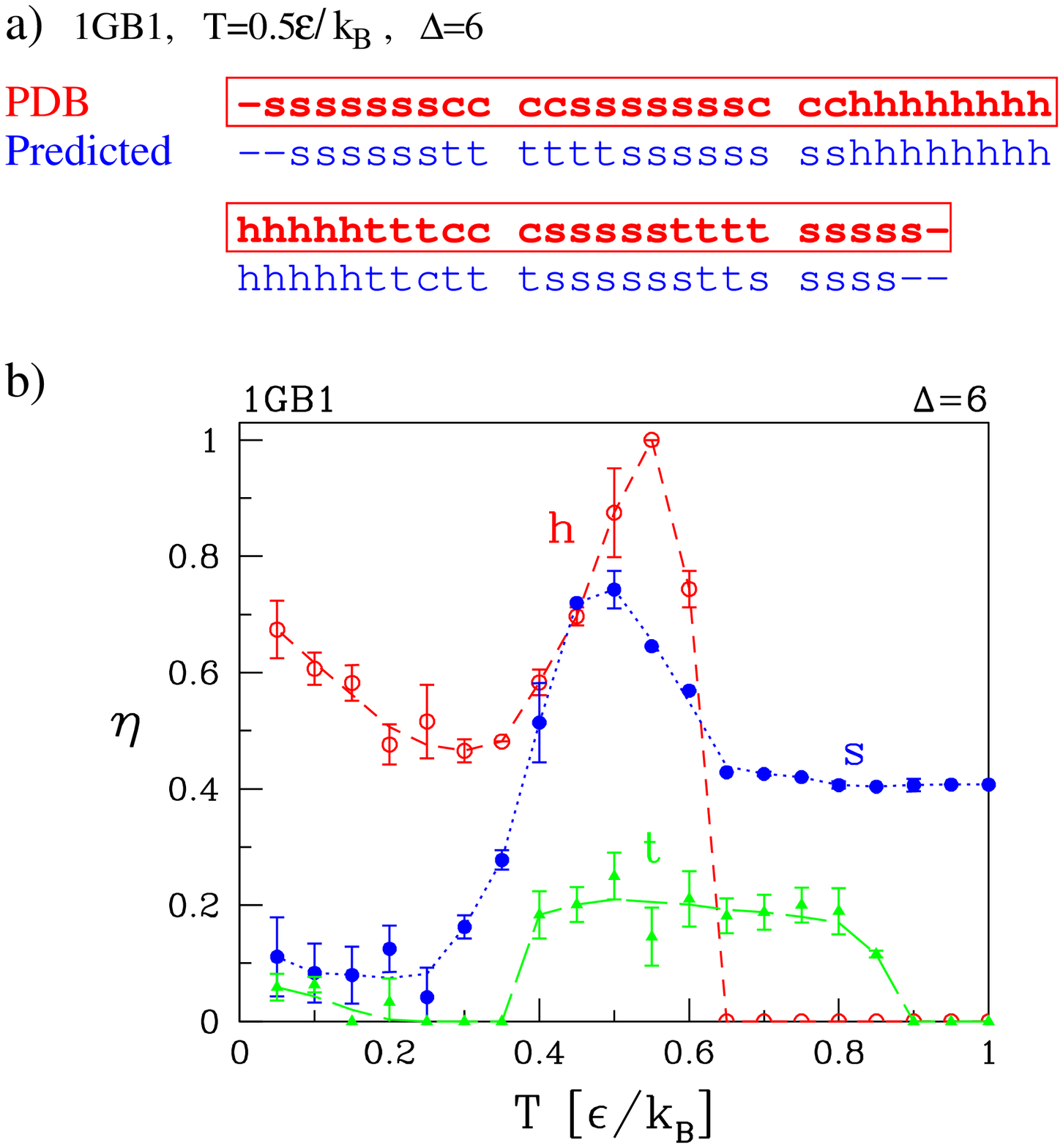}}
\caption{}
\label{sgb2}
\end{figure}

% FIGURE 3
\begin{figure}
\epsfxsize=3.2in
\centerline{\epsffile{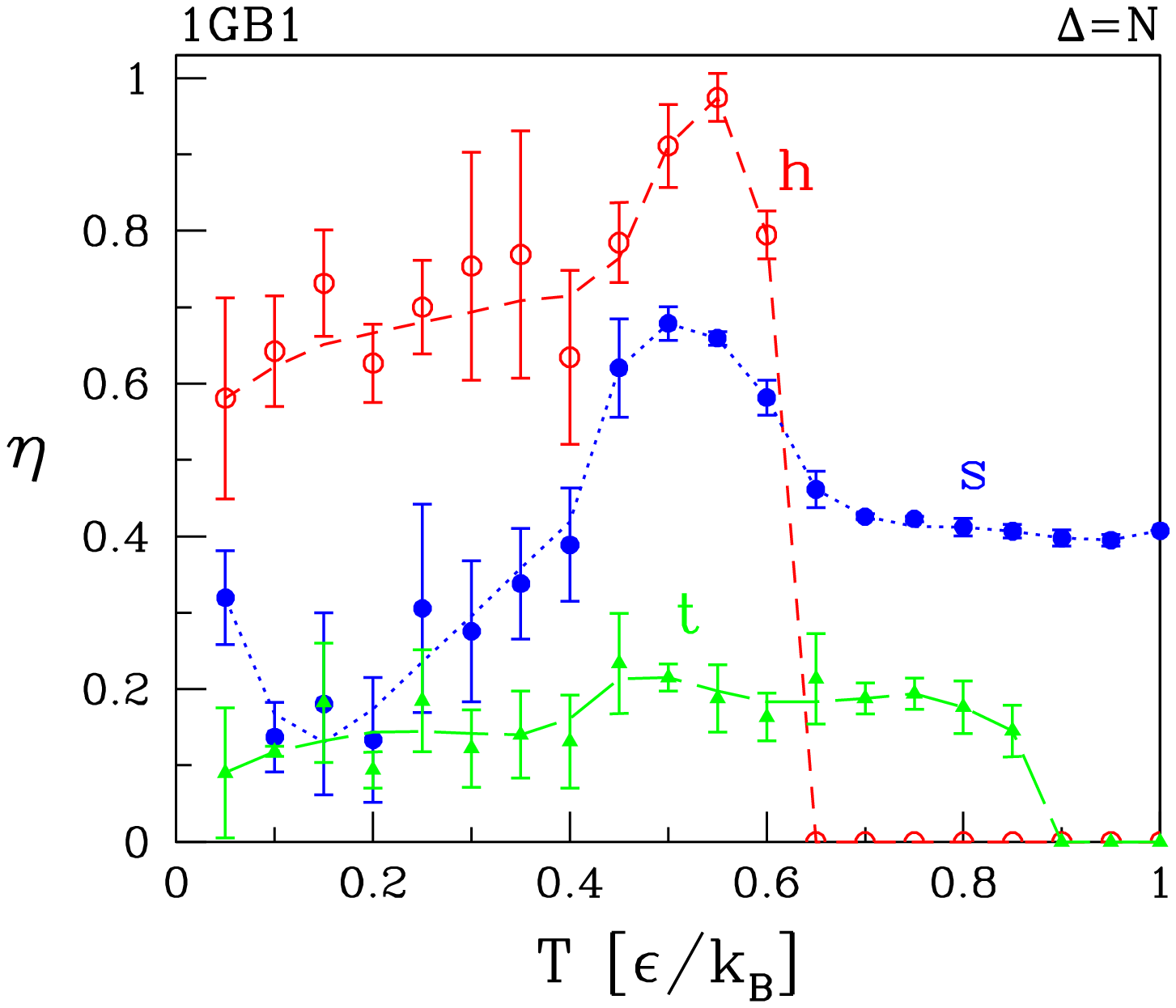}}
\caption{}
\label{agb2}
\end{figure}

% FIGURE 4
\begin{figure}
\epsfxsize=3.2in
\centerline{\epsffile{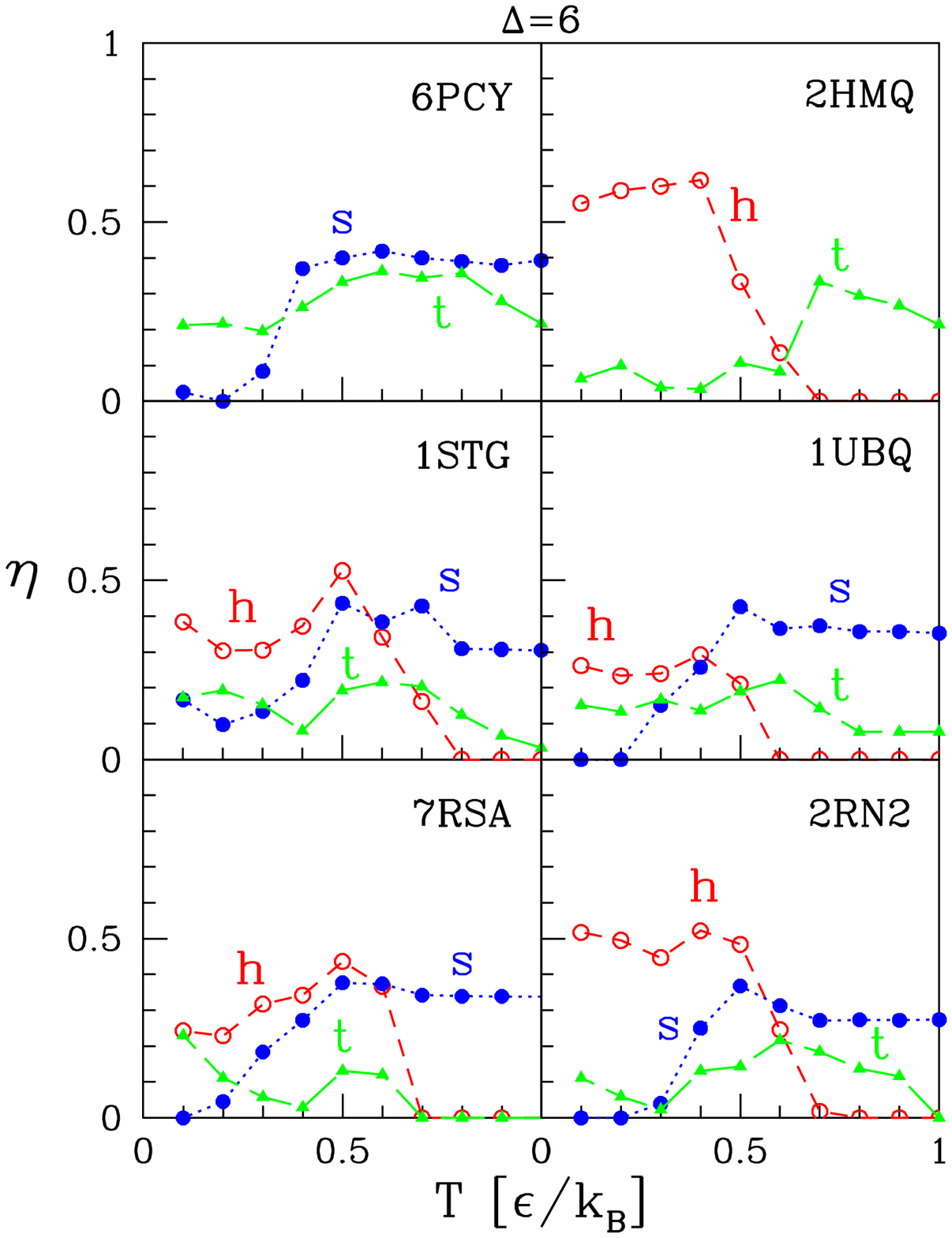}}
\caption{}
\label{eta}
\end{figure}

%\newpage
% FIGURE 5
\begin{figure}
\epsfxsize=3.2in
\centerline{\epsffile{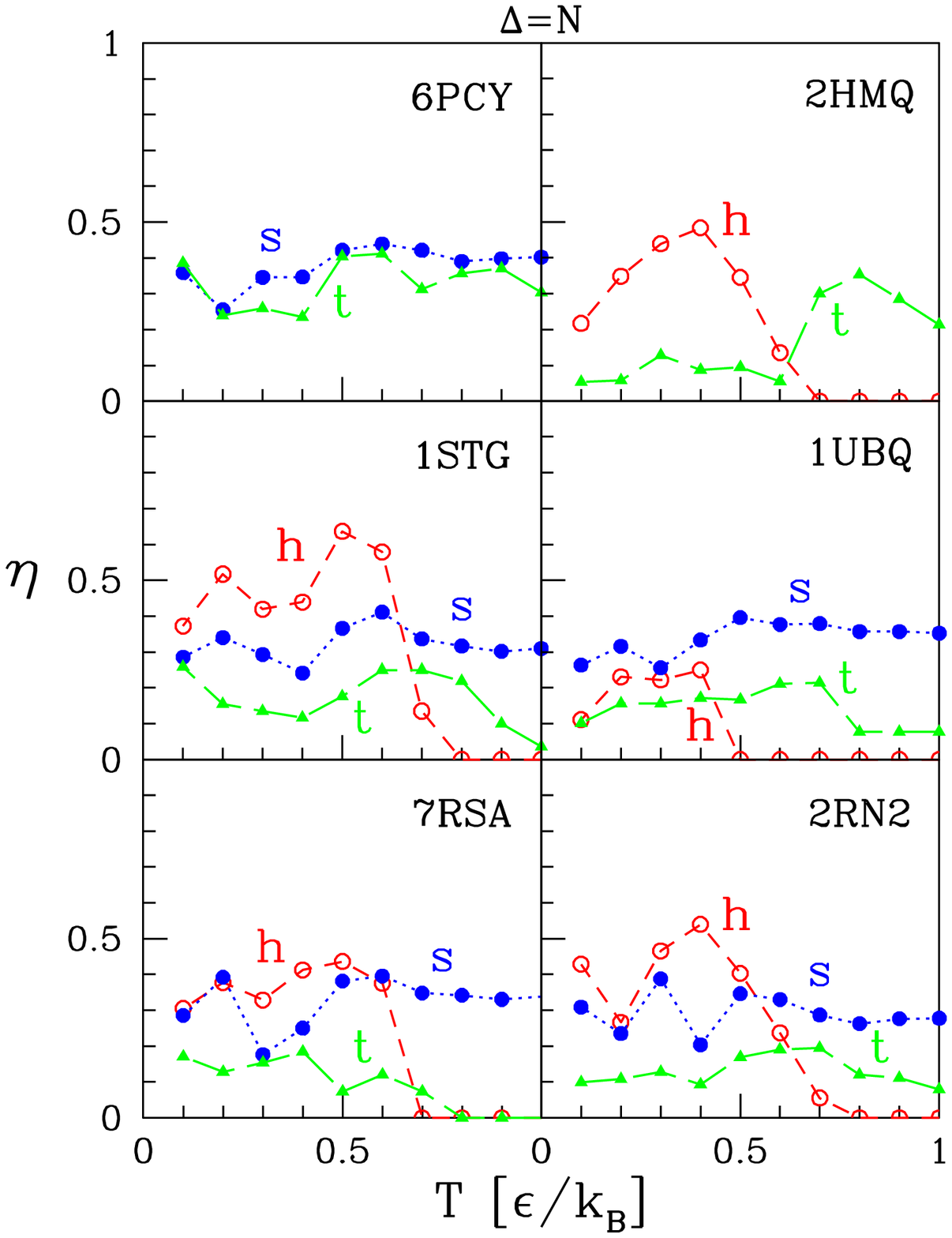}}
\caption{}
\label{apcy}
\end{figure}

% FIGURE 6
\begin{figure}
\epsfxsize=3.2in
\centerline{\epsffile{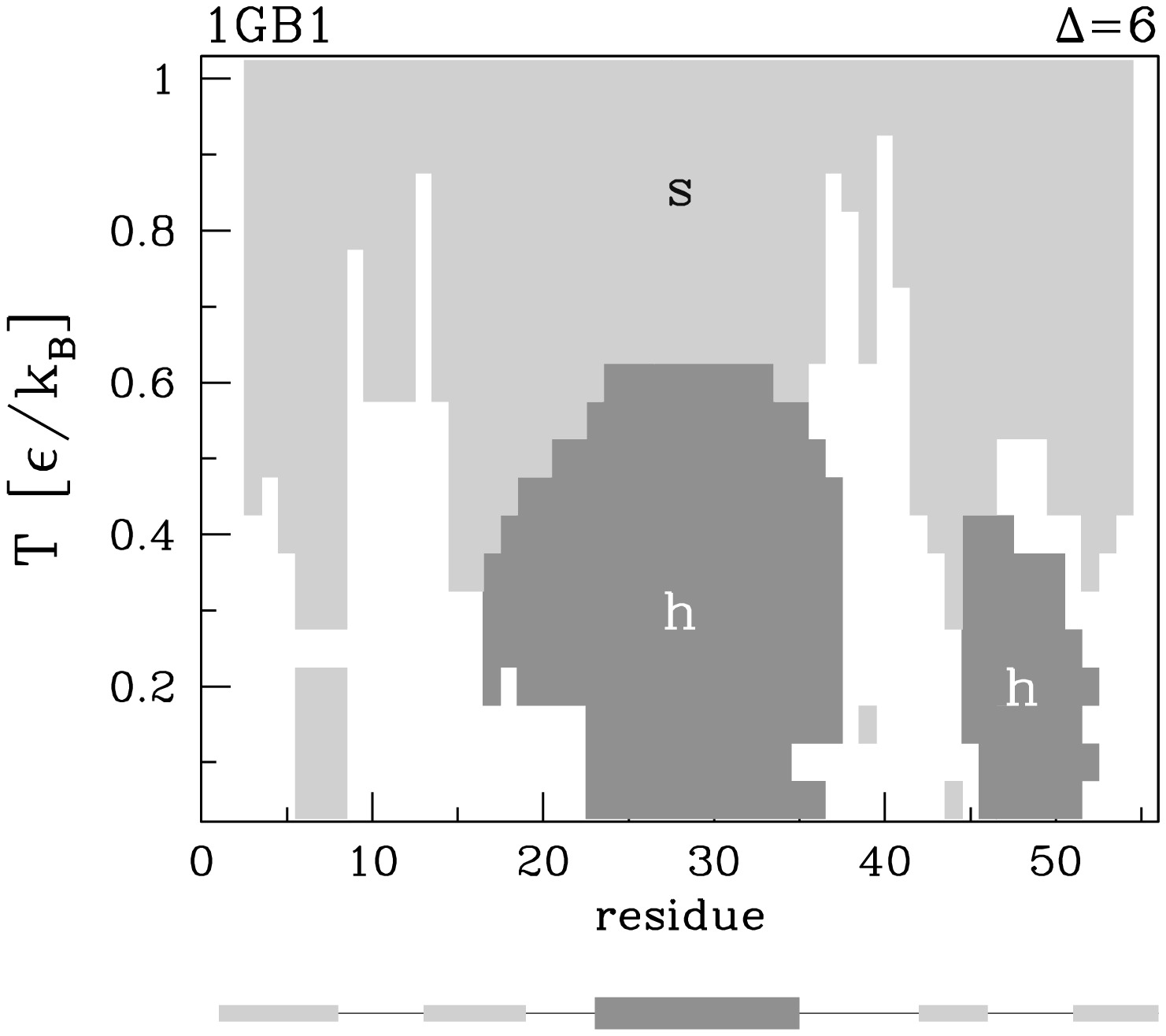}}
\caption{}
\label{phase}
\end{figure}

% FIGURE 7
\begin{figure}
\epsfxsize=3.2in
\centerline{\epsffile{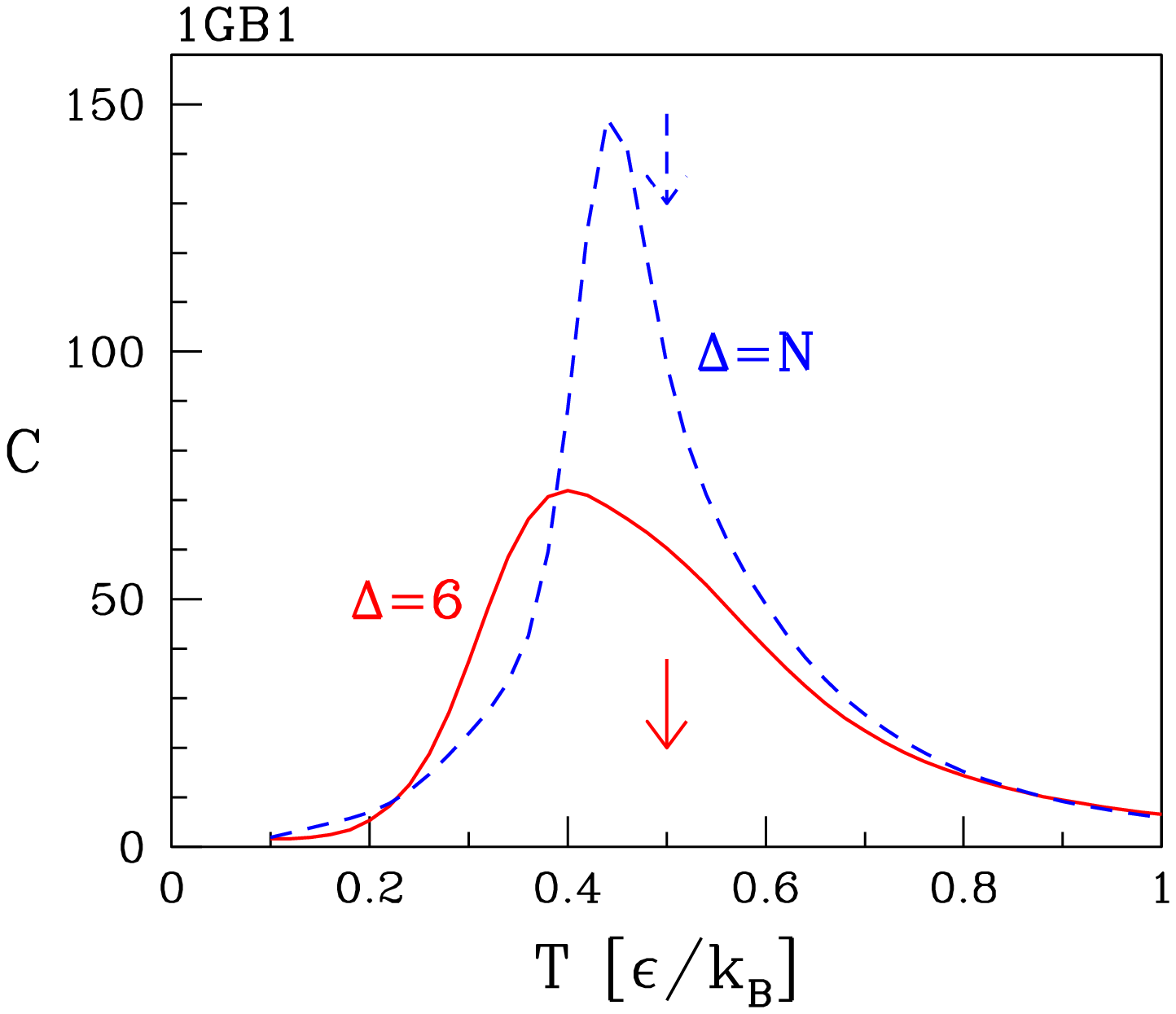}}
\caption{}
\label{cvgb1}
\end{figure}

% FIGURE 8
\begin{figure}
\epsfxsize=3.2in
\centerline{\epsffile{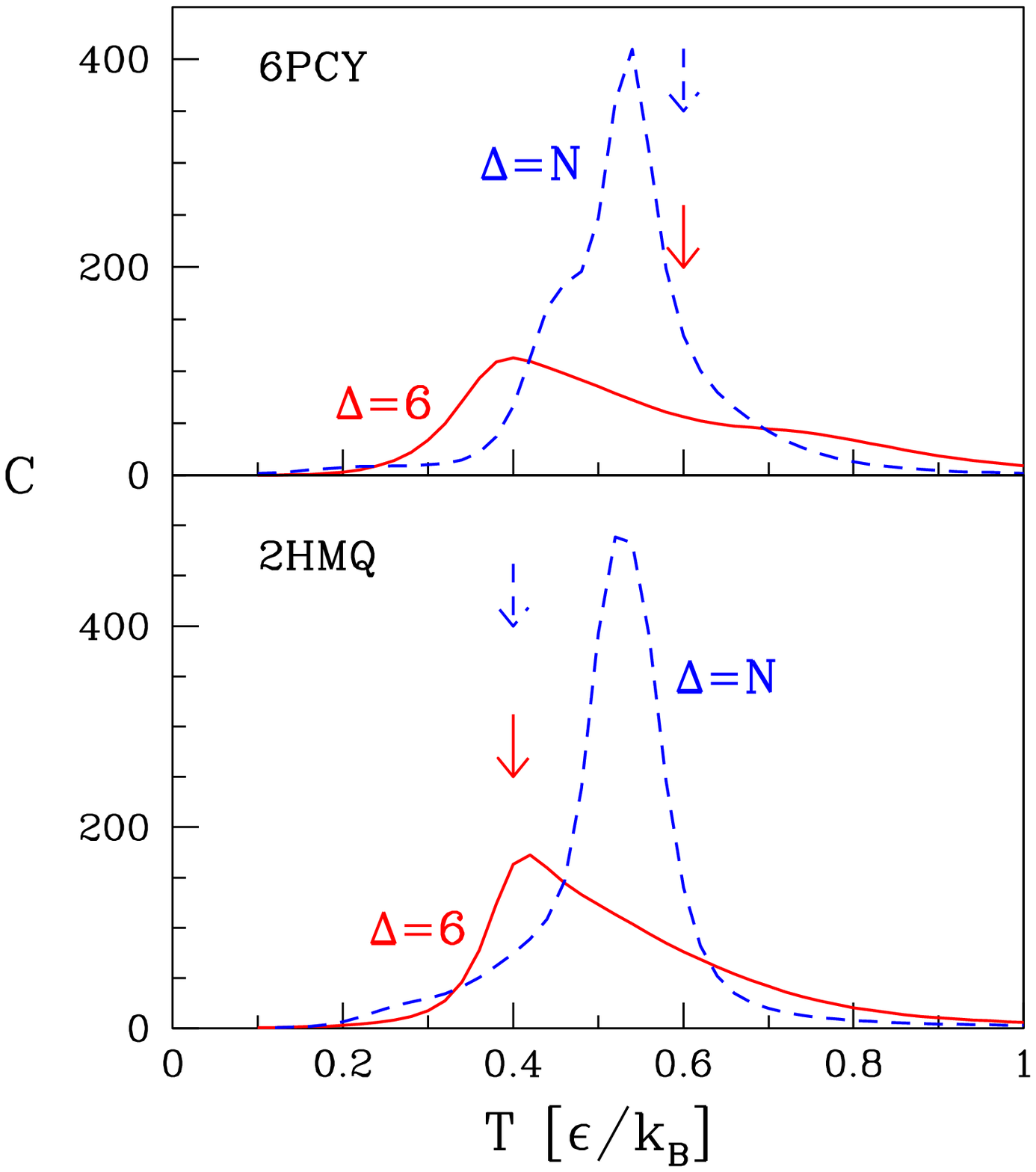}}
\caption{}
\end{figure}

% FIGURE 9
\begin{figure}
\epsfxsize=3.2in
\centerline{\epsffile{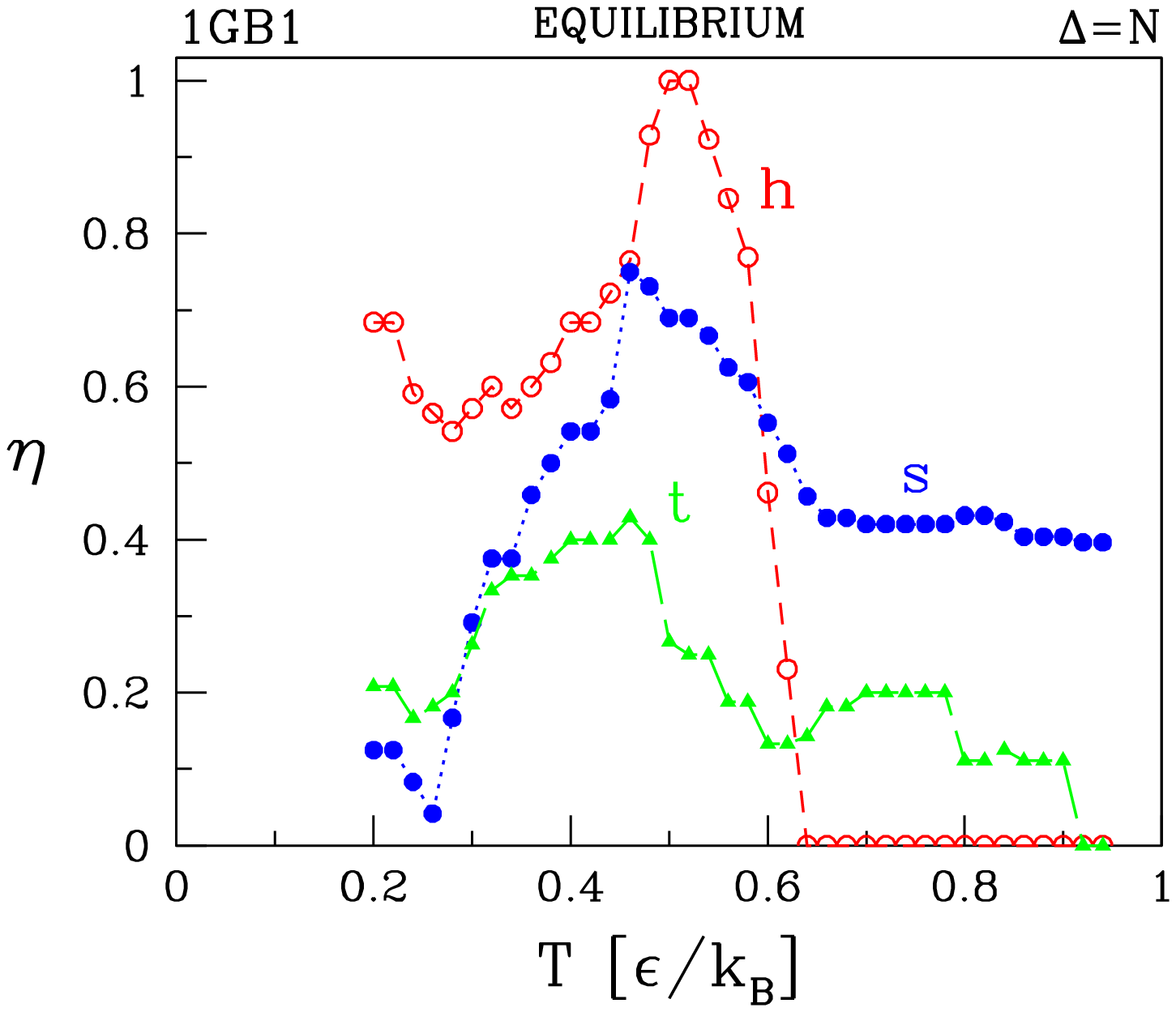}}
\caption{}
\label{agb3}
\end{figure}

\end{document}